\documentclass[aps, prl,reprint]{revtex4-1}
\usepackage{graphicx} 
\usepackage{epstopdf} 
\usepackage{color}

\begin{document}

\title{Ultrafast dynamics of neutral superexcited Oxygen: A direct measurement of the competition between autoionization and predissociation}

\author{Henry Timmers}
\email{timmers@physics.arizona.edu}
\author{Niranjan Shivaram}
\affiliation{Department of Physics, University of Arizona, Tucson, AZ, 85721 USA.}
\author{Arvinder Sandhu}
\email{sandhu@physics.arizona.edu}
\affiliation{Department of Physics, University of Arizona, Tucson, AZ, 85721 USA.}

\begin{abstract}
Using ultrafast extreme ultraviolet pulses, we performed a {\it direct} measurement of the relaxation dynamics of neutral superexcited states corresponding to the $nl\sigma_g(c^4\Sigma_u^-)$ Rydberg series of O$_2$. An XUV attosecond pulse train was used to create a temporally localized Rydberg wavepacket and the ensuing electronic and nuclear dynamics were probed using a time-delayed femtosecond near-infrared pulse. We investigated the competing predissociation and autoionization mechanisms in superexcited molecules and found that autoionization is dominant for the low $n$ Rydberg states. We measured an autoionization lifetime of $92\pm6$ fs and $180\pm10$ fs for $(5s,4d)\sigma_g$ and $(6s,5d)\sigma_g$ Rydberg state groups respectively. We also determine that the disputed neutral dissociation lifetime for the $\nu=0$ vibrational level of the Rydberg series is $1100\pm100$ fs.
\end{abstract}

\maketitle 

A pervasive theme in ultrafast science is the characterization and control of energy distributions in
elementary molecular processes. Ultrashort light pulses are used to excite and probe electronic
and nuclear wavepackets, whose evolution is fundamental to understanding many physical and chemical phenomena \cite{zewail1988}. However, until recently, time-resolved studies of wavepacket dynamics using light pulses in the infrared (IR), visible, and ultraviolet (UV) regime had predominantly been limited to low-lying excited states and femtosecond timescales \cite{Gessner2006}. 

Advances in photon technologies, specifically in the field of laser high-harmonic generation (HHG)\cite{rundquist1998,Paul2001}, have opened new avenues in the time-resolved studies of molecular dynamics. The efficient energy regime for HHG lies within the extreme ultraviolet (XUV) range from 10 eV to 100 eV, allowing for the excitation of inner-shell electrons to form highly excited atomic and molecular states.  Importantly, the attosecond to few-femtosecond nature of the HHG radiation forms a temporally localized excited state wavepacket whose dynamics can be followed in real-time using a time-delayed probe \cite{Drescher2002,Kelkensberg2009,Sansone2010,Singh2010,Goulielmakis2010}. Recently experiments have employed these attosecond XUV sources to study ultrafast fragmentation \cite{gagnon2007} and autoionization \cite{Sandhu2008,Cao2011} in highly excited molecular ions.

Apart from excited molecular ions, neutral molecules existing far above the ionization potential form an even more important class of molecular systems where the application of ultrafast XUV sources could provide new insights. Such neutral superexcited molecular states are often found as intermediates in chemical reactions initiated by high-energy photons (e.g., solar radiation) and play an important role in the chemistry of planetary atmospheres \cite{Wayne1991}. The superexcited molecules quickly relax from their non-equilibrium state through multiple competing decay mechanisms including autoionization, fluorescence, and dissociation into neutral fragments \cite{Hatano2003,Ukai,Li2004,Strasser2008}. Synchrotron light sources have  traditionally been employed for investigating these processes. While these sources offer excellent spectral resolution, they lack the time resolution required for the real-time study of ultrafast processes in superexcited molecules. In contrast, an attosecond pump-probe experiment can resolve dynamics on the few femtosecond timescale.

\begin{figure}
   \centering
   \includegraphics [ width=.7\linewidth ] {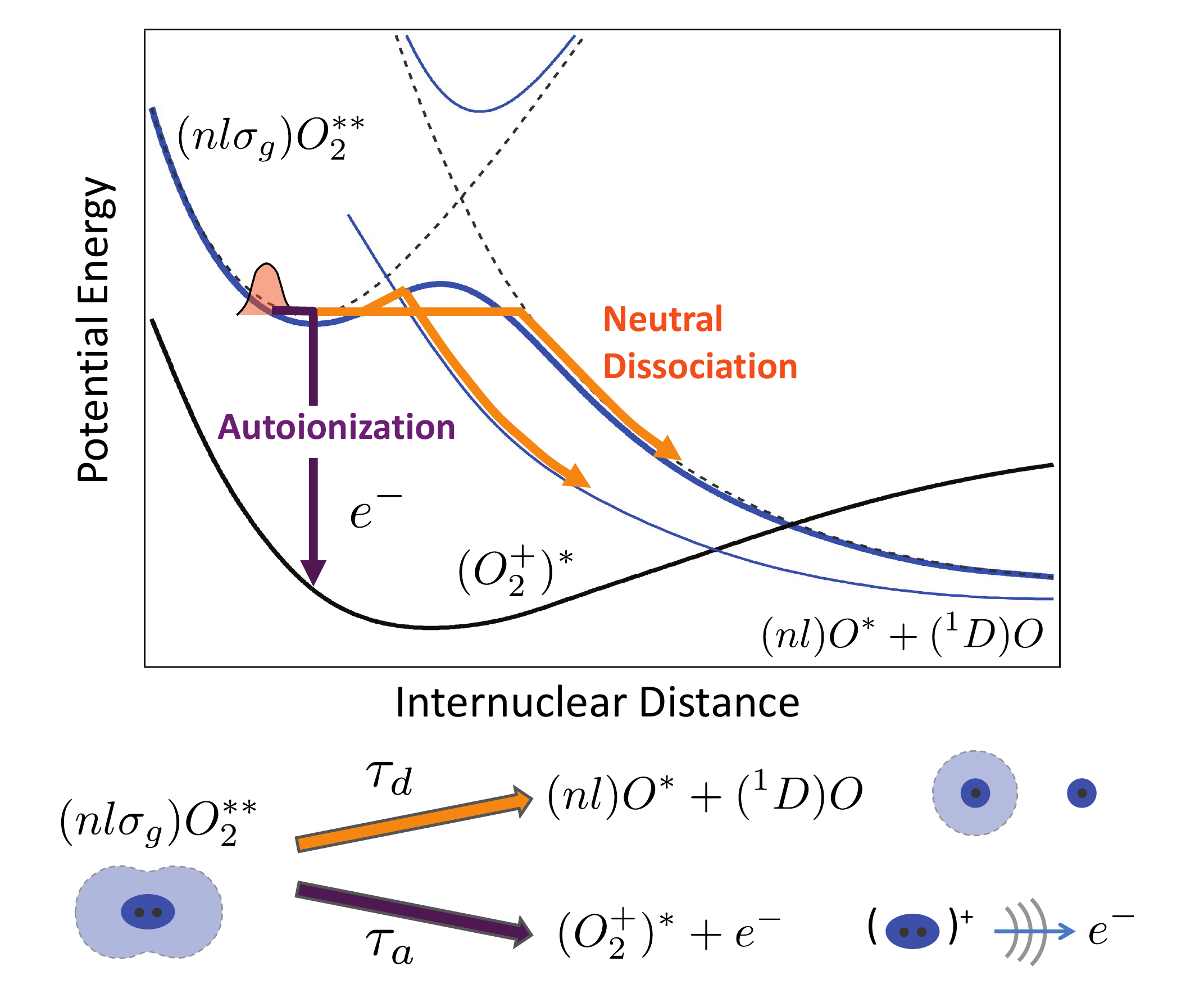}
   \caption{An ultrashort XUV pulse excites a $2s\sigma_u$ electron to form the $nl\sigma_g(c^4\Sigma_u^-)$ Rydberg state of O$_2$ (blue curve). Fast predissociation of these highly excited molecules into neutral fragments can occur through non-adiabatic interactions. Since these neutral states are embedded in an ionization continuum, autoionization can also occur through spontaneous ejection of an electron. The neutral dissociation and autoionization thus form competing decay mechanisms.}
   \label{Fig1Ryd}
\end{figure}

In this work, we focus on the dynamics of the superexcited oxygen molecule, specifically the $nl\sigma_g$ family of neutral Rydberg states converging to the $c^4\Sigma_u^-$ state of O$_2^+$. These superexcited states are formed through the direct excitation of an inner-shell $2s\sigma_u$ electron to states that lie greater than 10 eV above the ionization potential of O$_2$. The generic $nl\sigma_g$ Rydberg state curve resulting from the interaction between two electronic configurations of O$_2$ is shown in Fig. \ref{Fig1Ryd}. The quasi-bound vibrational population in this state can predissociate very rapidly through tunneling and/or non-adiabatic mechanisms, such as spin-orbit interaction and rotational coupling\cite{Frasinski}. The neutral excited oxygen atoms formed in this dissociation play a important role in the dynamics of ozone formation and decay in Earth's upper atmosphere \cite{Wayne1991}. Further, the superexcited states are also embedded in the ionization continuum where autoionization forms another important relaxation channel. In this process, an electron is spontaneously ejected from the superexcited neutral molecule resulting in a molecular ion (Fig. \ref{Fig1Ryd}). The superexcited O$_2$ system thus presents a unique opportunity to study the competition between neutral dissociation and autoionization and explore the non-adiabatic effects which dominate the dynamics of superexcited molecules. However, despite the significant interest and the fundamental importance of these dynamics, crucial questions remain unanswered. 

Prior synchrotron measurements and theoretical studies\cite{Tanaka, Frasinski, Akahori, Evans, Hikosaka, Ehresmann, Reddish, Demekhin1} have explored the fragmentation of $c^4\Sigma_u^-$ ionic core and associated high $n$ Rydberg states. These studies indicate that the dynamical evolution of high $n$ Rydberg levels mimics the $c^4\Sigma_u^-$ ion-core dissociation and neutral dissociation is the dominant relaxation mechanism. Of the two vibrational levels ($\nu=0,1$) supported by these quasibound states, consensus has been reached that the predissociation lifetime for the $\nu=1$ vibrational level is $\sim70$fs\cite{Hikosaka, Evans, Reddish, Ehresmann, Demekhin1}. However, the dissociation lifetimes of the $\nu=0$ vibrational level still differ by orders of magnitude between different studies \cite{Hikosaka, Evans,Demekhin1}. Furthermore, in contrast to the high $n$ Rydberg states, the electron correlation effects in the low $n$ Rydberg states (e.g. $n=4,5,6$) lead to a scenario in which autoionization can become a competing and even prominent decay mechanism. To date, the autoionization rates for low $n$ Rydberg states have not been measured. The bottlenecks in accurate frequency-domain determination of the predissociation and autoionization lifetimes stem from the complications associated with the fitting of overlapping resonances and asymmetric lineshapes. Here, we report a time-domain pump-probe study that allows a direct determination of the autoionization lifetimes for low $n$ Rydberg states and simultaneously provides a measurement of the disputed neutral predissociation lifetime for the $\nu=0$ level.

In our experiment, a 1.5 mJ near-infrared (IR) pulse with a temporal width of 45 fs was divided into a pump and probe path.  The pump pulse was focused into a hollow-core waveguide filled with Xenon gas to create an XUV attosecond pulse train (APT) consisting primarily of the 13th and 15th harmonics. The temporal duration of the APT envelope was approximately 3 fs. The XUV pulse was then focused into an effusive gas jet of molecular oxygen to create a temporally localized wavepacket corresponding to Rydberg $nl\sigma_g$ states. A probe IR pulse with a peak intensity of 4 TW/cm$^2$ was used to ionize the neutral excited wavepacket resulting in the formation of the continuum molecular ion, ($c^4\Sigma_u^-$)O$_2^+$. This state served as our final product channel as it leads to formation of easily observable O$^+$ ions. We used a velocity map imaging (VMI) detector to measure the kinetic energy and yield of O$^+$ ions in two dissociation limits as a function of pump-probe time delay. The time-zero of the pump-probe scan was calibrated to an accuracy of $\pm2$fs using a crosscorrelation measurement in the form of two-photon (XUV+IR) Helium photoionization.

\begin{figure}
   \centering
   \includegraphics[width=0.95\linewidth]{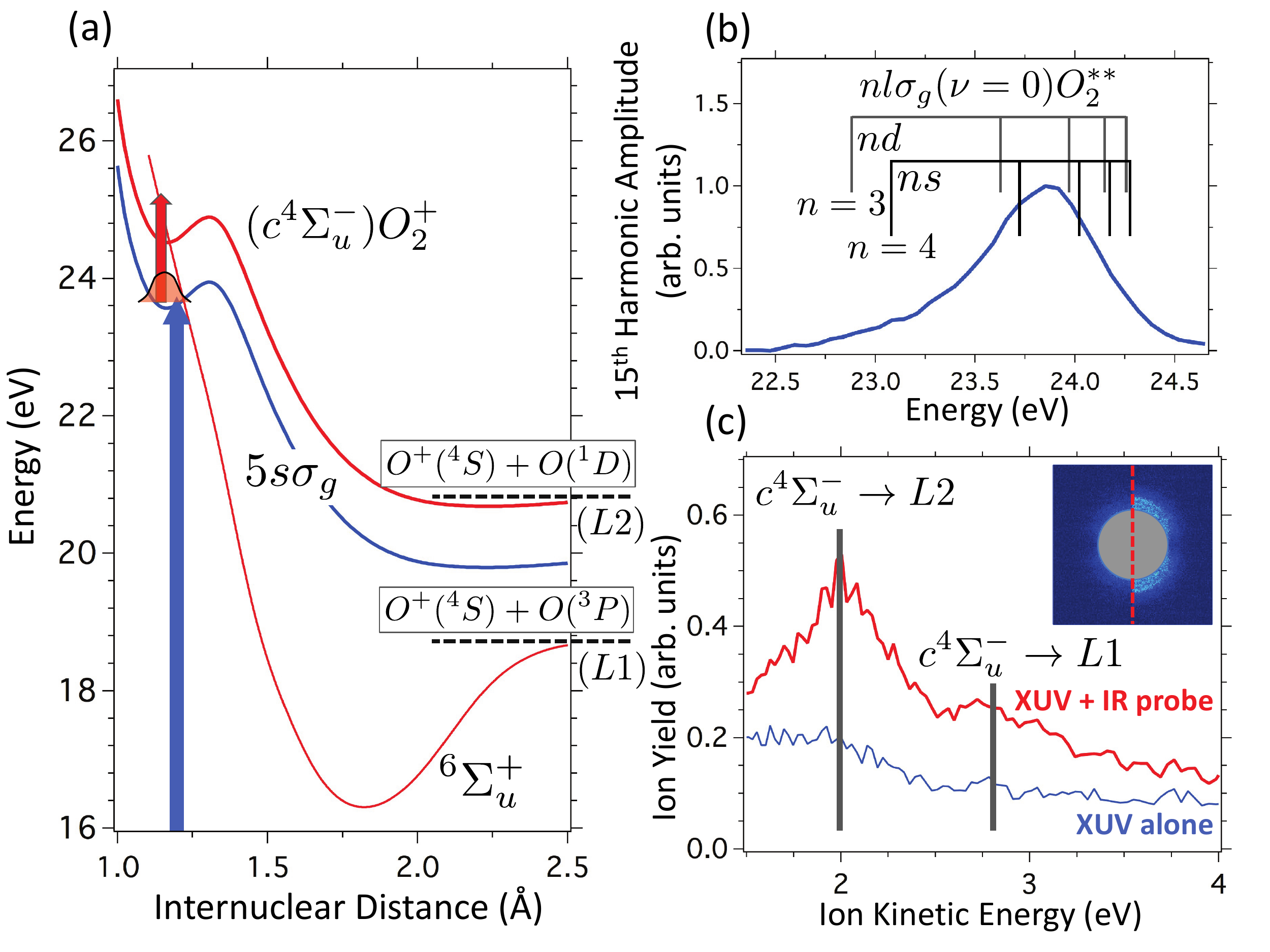}
   \caption{(a) A schematic representation of the relaxation lifetime measurement. The $5s\sigma_g$ Rydberg state is populated by the XUV APT and decays rapidly due to autoionization and predissociation. We interrupt this decay process by removing the Rydberg electron with a time-delayed IR pulse to populate the $(c^4\Sigma_u^-)$O$_2^+$ state and produce measurable ions with well-defined kinetic energies corresponding to $c^4\Sigma_u^-\rightarrow$L1 and $c^4\Sigma_u^-\rightarrow$L2 channels. (b) The XUV excitation spectrum for 15th harmonic used in the lifetime measurements overlaid with the $n(s,d)\sigma_g$ Rydberg assignments. (c) The ion kinetic energy spectrum with the XUV pulse alone (lower curve, blue) as well as in the simultaneous presence of the XUV + IR probing pulse (upper curve, red). Inset shows the full velocity map of ions with (right half) and without(left half) the IR pulse.}
   \label{Fig2Probing}
\end{figure}

For the measurement of the relaxation lifetime for superexcited Rydberg states, we tune the spectrum of our XUV pulse such that the peak of the 15th harmonic is resonant with the $n(s,d)\sigma_g$ Rydberg states (Fig. \ref{Fig2Probing}(a)). The $nl\sigma_g(c^4\Sigma_u^-)$ Rydberg state assignments for the $\nu=0$ vibrational level obtained from prior synchrotron work \cite{Hikosaka} are shown in Fig. \ref{Fig2Probing}(b) along with our excitation spectrum. The excitation of the $\nu=1$ vibrational manifold is expected to be factor of three times weaker than $\nu=0$ \cite{Evans}. 

The temporally localized population in the Rydberg excited state decays very rapidly due to the strong electron correlation effects and the quasi-bound nature of the potential energy curve. We interrupt this decay with the IR probe pulse to ionize the molecule, leading to the formation of  $(c^4\Sigma_u^-)$O$_2^+$ which fragments along the known pathways \cite{Johnsson2008} $c^4\Sigma_u^-\rightarrow$L1 and $c^4\Sigma_u^-\rightarrow$L2. These two paths lead to O$^+$ ions with kinetic energies around 2.9 eV and 1.9 eV respectively. In Fig. \ref{Fig2Probing}(c) we plot the kinetic energy spectrum of $O^+$. The lower curve represents the background in photo-ion counts due to XUV alone.  The upper curve shows the photo-ion counts in the simultaneous presence of both the XUV and IR fields. Clearly, the ion counts in both the L1 and L2 channels are enhanced greatly by the action of the IR probe pulse. 

The $c^4\Sigma_u^-\rightarrow$L1 channel near 2.9eV is of particular interest in the current study since it arises purely from the predissociation of the $\nu=0$ vibrational level\cite{Frasinski, Akahori}. Since the excitation from Rydberg states to the ionic-core is dominated by $\Delta\nu=0$ transitions, the ion yield measured in the L1 limit originates exclusively from the $\nu=0$ population of the Rydberg states. Thus, the measurement of $c^4\Sigma_u^-\rightarrow$L1 ion yield as a function of pump-probe time delay  provides us an avenue to measure both the controversial $\nu=0$ predissociation rate and the unknown autoionization rate.

We obtain a kinetic model of the relaxation of $\nu=0$ level by considering the decay of a single Rydberg state due to the competition between autoionization and neutral dissociation. A straightforward derivation begining with the relevant rate equations\cite{Supplement} gives the delay dependence of the ion-yield in the $c^4\Sigma_u^-\rightarrow$L1 channel as
\begin{equation}
Y_{L1}(P_o,\tau_a,\tau_d;t)=\alpha P_o \left[\frac{\tau_d}{\tau_a+\tau_d}e^{-t(1/\tau_a+1/\tau_d)}+\frac{\tau_a}{\tau_a+\tau_d}\right],
\label{L1Eq}
\end{equation}
where $P_o$ is the initial population in the $\nu=0$ Rydberg vibrational level,  $\tau_a$ is the autoionization lifetime, $\tau_d$ is the predissociation lifetime, and $\alpha$ is the branching ratio for $\nu=0$ population into the L1 channel.

\begin{figure}
   \centering
   \includegraphics [ width=0.7\linewidth ] {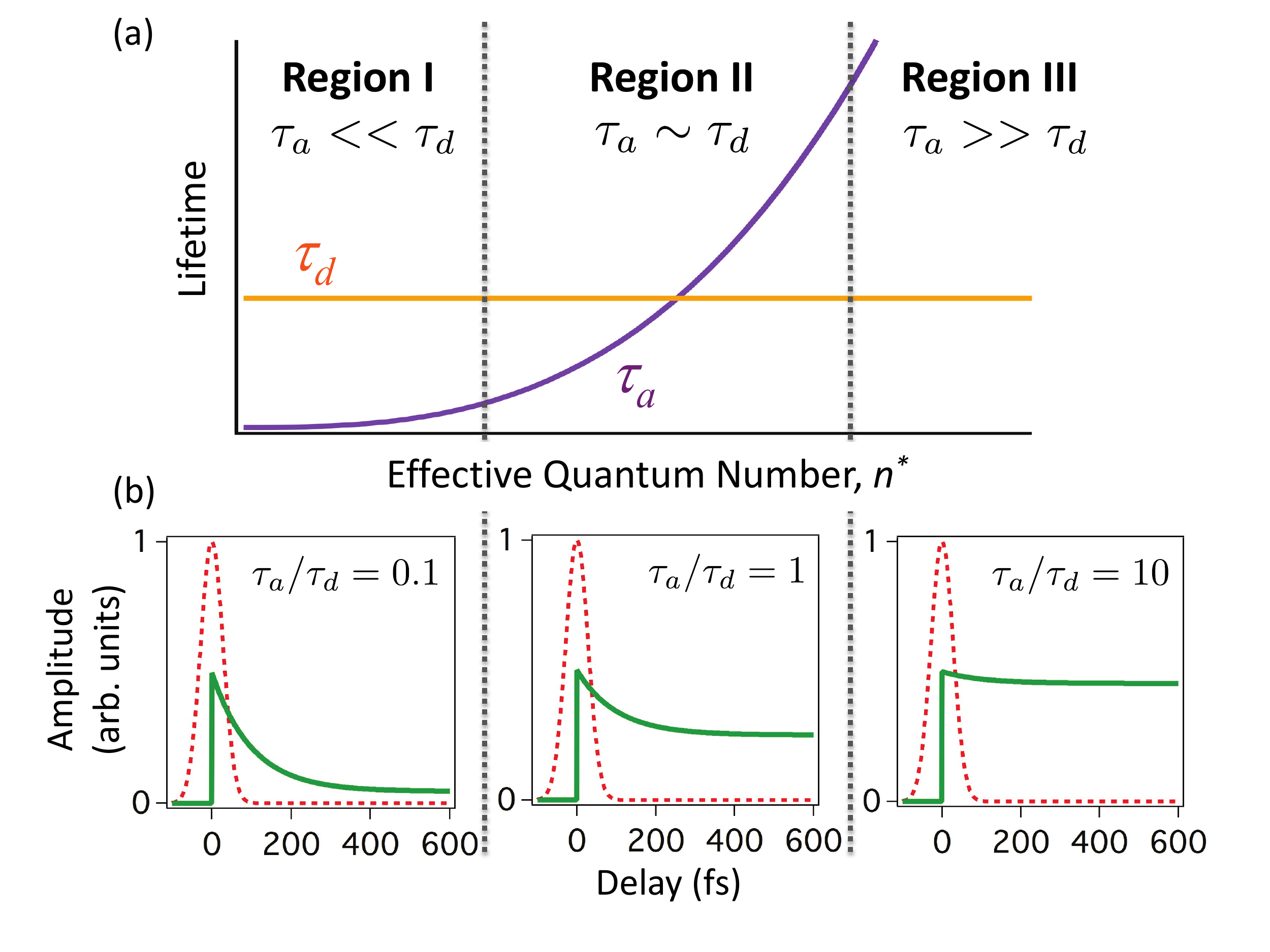}
   \caption{(a) A sketch showing $\tau_a$ and $\tau_o$ scaling with effective quantum number $n^*$.  (b) The subplots illustrate the simulated delay dependence of the ion yield in three regimes (green curves) which differ in the relative contribution of predissociation and autoionization mechanisms for a fixed total decay lifetime of 100fs. Red curves depict the Gaussian pump-probe cross-correlation profile representing additional direct XUV+IR excitation processes.}
   \label{Fig3Model}
\end{figure} 

This equation can be understood in the context of the multichannel quantum defect theory for Rydberg states in molecules \cite{Fano1970} where the effective quantum number $(n^*)$ incorporates the effect of core interactions. Using this formalism, it has been established that both the oscillator strength for excitation to a $nl\sigma_g$ Rydberg state, $f_{nl}$, and the autoionization rate from that Rydberg state scale as $(n^*)^{-3}$. However, in the core-ion approximation\cite{Hikosaka}, the predissociation rate is independent of the effective quantum number of the Rydberg state. The contrasting $n^*$ dependence of autoionization and neutral dissociation lifetimes is shown in Fig. \ref{Fig3Model} (a). Using Eq. \ref{L1Eq}, the variation of ion-yield with time delay can be simulated in three regimes. For low $n^*$ states (Region I, Fig. \ref{Fig3Model} (b)), autoionization of the $\nu=0$ population from the potential well is the dominant mechanism, manifesting itself as an exponential decay to a near zero baseline in the $c^4\Sigma_u^-\rightarrow$L1 channel ion yield (green curve). For high $n^*$ states (Region III, Fig. \ref{Fig3Model} (b)), neutral dissociation is the dominant mechanism. However, as the neutral fragments can be photoionized in the dissociation limit there is no net loss of experimental ion counts. The delay-dependent ion signal therefore exhibits a near flat, dc response. When both mechanisms have comparable rates (Region II, Fig. \ref{Fig3Model} (b)), the exponential decay terminates into an appreciable dc baseline. In general, the amplitude of the exponential term relative to the dc baseline in Eq. \ref{L1Eq} elucidates the competition between autoionization and neutral dissociation. This allows us to decouple the two decay mechanisms and measure the contribution from each process. 

Before we quantitatively deduce the autoionization and predissociation lifetimes, we note that Eq. \ref{L1Eq} refers to ion-yield obtained from a single Rydberg level. From Fig. \ref{Fig2Probing}(b), we observe that the Rydberg states $5s\sigma_g$, $4d\sigma_g$, $6s\sigma_g$ and $5d\sigma_g$ are predominantly populated by the 15${th}$ harmonic. Based on the similarity between the energies and effective quantum numbers, it is convenient to classify these Rydberg states \cite{Demekhin2} into two groups i.e. $(5s,4d)\sigma_g$ and $(6s,5d)\sigma_g$. Since the harmonic field strength is nearly identical for both groups, the oscillator strength for each group determines its excitation probability. Defining $r=(n^*_{5s,4d})^3/(n^*_{6s,5d})^3$ as the scaling factor for the oscillator strengths and autoionization lifetimes between consecutive Rydberg states, we replace $P_o$ with $rP_o$ and $\tau_a$  with $\tau_a/r$ in Eq. \ref{L1Eq} to obtain the yield of the second Rydberg group. The time dependence of total yield can thus be written as  $Y_{L1}^{total}(t) = Y_{L1}^{5s,4d}(P_o,\tau_a,\tau_d;t)+ Y_{L1}^{6s,5d}(rP_o,\tau_a/r,\tau_d;t)$. We invoke Rydberg assignments reported by Hikosaka \textit{et al.} \cite{Hikosaka} to fix the constant $r=0.515$. Lastly, we note that while we do not include the effects of the higher lying Rydberg states, the population of these states is substantially smaller ($\le 10\%$) and should not affect our analysis.

\begin{figure}
   \centering
   \includegraphics [ width=0.9\linewidth ] {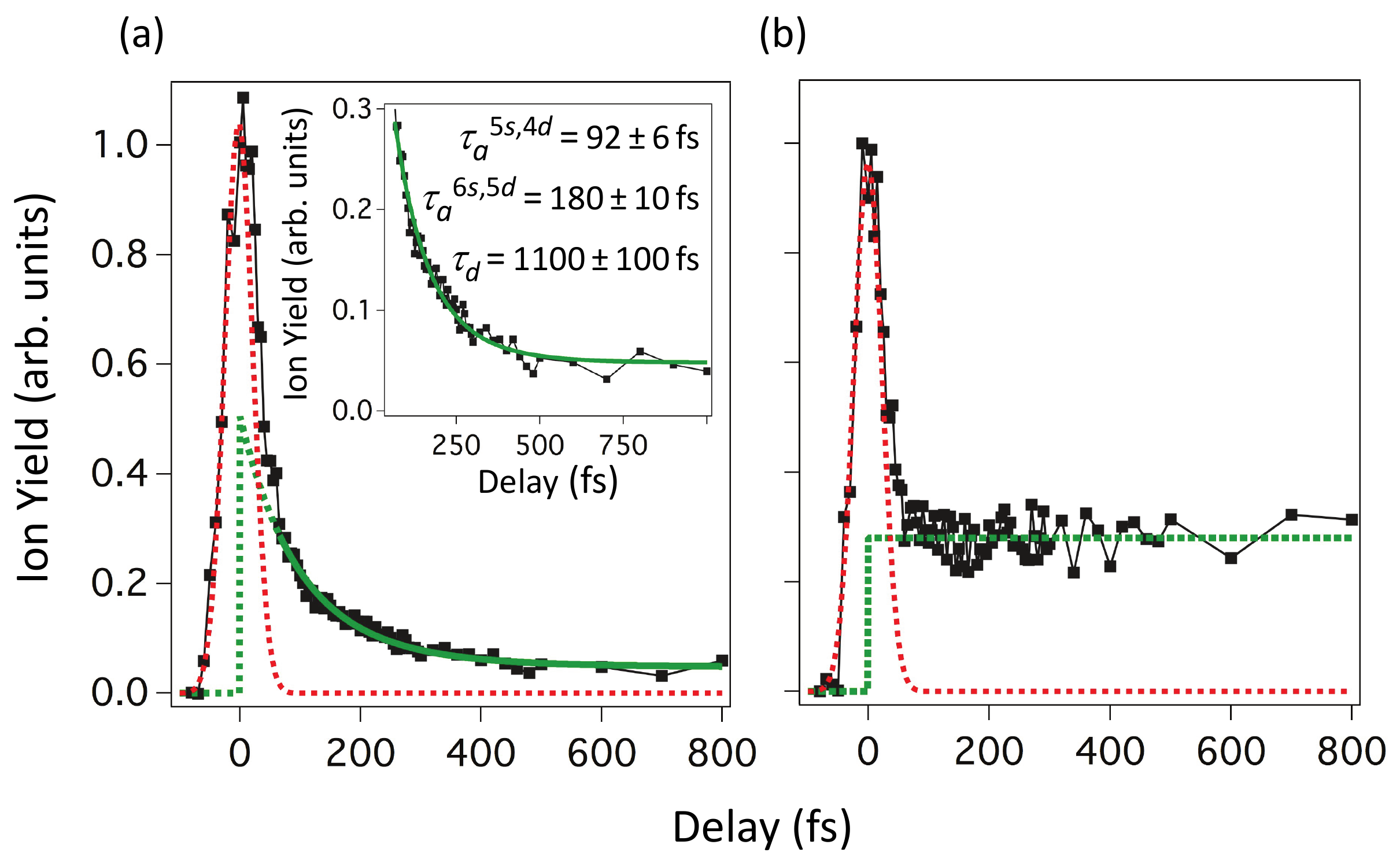}
   \caption{(a) Experimental ion yield measured in the $c^4\Sigma_u^-\rightarrow$L1 channel. The fit obtained for time delays $\ge$70 fs using the model discussed in the text is shown as a solid green line (a zoomed in view of the fitting region is shown in the inset). (b) Experimental ion yield measured in the $c^4\Sigma_u^-\rightarrow$L1 channel when the harmonic spectrum is tuned to be resonant with the high $n$ Rydberg states. In this region, neutral dissociation dominates resulting in the expected flat delay-dependence of the ion yield.} 
   \label{Fig4Exp}
\end{figure}

The delay dependent ion yield obtained in the $c^4\Sigma_u^-\rightarrow$L1 channel is plotted in Fig. \ref{Fig4Exp}(a). The dashed red curve depicts the extent of the Gaussian cross-correlation between pump and probe pulses and represents the occurrence of additional processes due to the simultaneous presence of XUV and IR photons. We confine our analysis to time delays greater than 70 fs where temporal overlap effects can be ignored. We fit the data beyond 70 fs to the fitting function $Y_{L1}^{total}(t)$ which has three free parameters $P_o$,$\tau_a$,$\tau_d$, with $P_o$ being an overall multiplier. While the fitting expression consists of two exponential terms and a dc term, the amplitudes, decay rates and the dc baseline are all inter-related through the $\tau_a$ and $\tau_d$ parameters. Thus, we have a tightly constrained system and the $Y_{L1}^{total}(t)$ fit converges to unique values for $\tau_{a}$ and $\tau_{d}$. The fit shown by the solid green curve in Fig. \ref{Fig4Exp}(a) yields an autoionization lifetime of $\tau_{a}^{5s,4d}=92\pm6$ fs and a neutral dissociation lifetime of $\tau_d=1100\pm100$ fs. Using the scaling factor $r=0.515$, the autoionization lifetime of the second Rydberg group is given by $\tau_{a}^{6s,5d}=180\pm10$ fs. These results show that autoionization is the dominant relaxation process for low to mid $n$ Rydberg states. The same understanding can be obtained through a simple comparison of the shape of our experimental curve (Fig. \ref{Fig4Exp} (a)) to the simulated plots shown in Fig. \ref{Fig3Model}(b).

We can look at the other extreme by tuning our harmonic spectrum to be on resonance with the high $n$ Rydberg states near the ionic continuum.  Since $\tau_a$ grows as $(n^*)^3$, we expect predissociation to dominate, resulting in an ion yield that exhibits no dependence on IR pump-probe delay (Region III, Fig. \ref{Fig3Model} (b)). This is demonstrated experimentally in Fig. \ref{Fig4Exp} (b), thus confirming the validity of our analysis. 

To the best of our knowledge, there are no prior experimental measurements of autoionization lifetimes for these superexcited Rydberg states. In a recent theoretical effort\cite{Ehresmann,Demekhin2}, lifetimes of $\tau_a^{5s}=90.0$ fs ($\tau_a^{4d}=92.7$ fs) and $\tau_a^{6s}=178$ fs ($\tau_a^{5d}=183$ fs) have been reported, which are in a good agreement with our experimental observations. We note this agreement also justifies the Rydberg state groupings that we had introduced in our model. The measured value for predissociation lifetime $\tau_d$ is also in agreement with the experimental lower bounds measured by Hikosaka \textit{et al.} \cite{Hikosaka} and Padmanabhan \textit{et al.} \cite{Reddish} who argued $\tau_d>600$ fs and $\tau_d\geq1$ ps respectively.

In summary, we were able to disentangle and measure the competing autoionizaton and neutral dissociation lifetimes of the $(5s,4d)\sigma_g$ and $(6s,5d)\sigma_g$ Rydberg states of O$_2^{**}$, formed through the XUV excitation of a $2s\sigma_{u}$ electron. This represents the first direct measurement of the ultrafast autoionization and predissociation dynamics in a neutral superexcited molecule. This topic has been of significant interest to the synchrotron community over last three decades. The knowledge of neutral dissociation and autoionization lifetimes can be used as a stepping stone to explore the relative importance of various spin-orbit interaction terms and tunneling processes in predissociation dynamics. Thus, the table-top HHG based XUV time-domain technique used here represents a significant step forward in the understanding of non-adiabatic relaxation mechanisms. In general, the coupling between electronic and nuclear degrees of freedom forms an interesting problem where direct time domain techniques could be very illuminating\cite{Zhou,Allison}. An important facet of the time-domain experimentation is the possibility of implementing real-time control, wherein a light pulse can be used to modify or change the course of the photo-chemical reaction in the transient phase.

We gratefully acknowledge the support from NSF grant PHY-0955274 and TRIF Imaging Fellowship. We also thank Dr. Xiao-Min Tong for useful discussions.

\bibliographystyle{apsrev4-1}
\bibliography{O2Refs}

\end{document}